# Procedural City Modeling

Thomas Lechner[1]    Ben Watson[1]    Uri Wilensky[1]    Martin Felsen[2]

Northwestern University[1]    Illinois Institute of Technology[2]

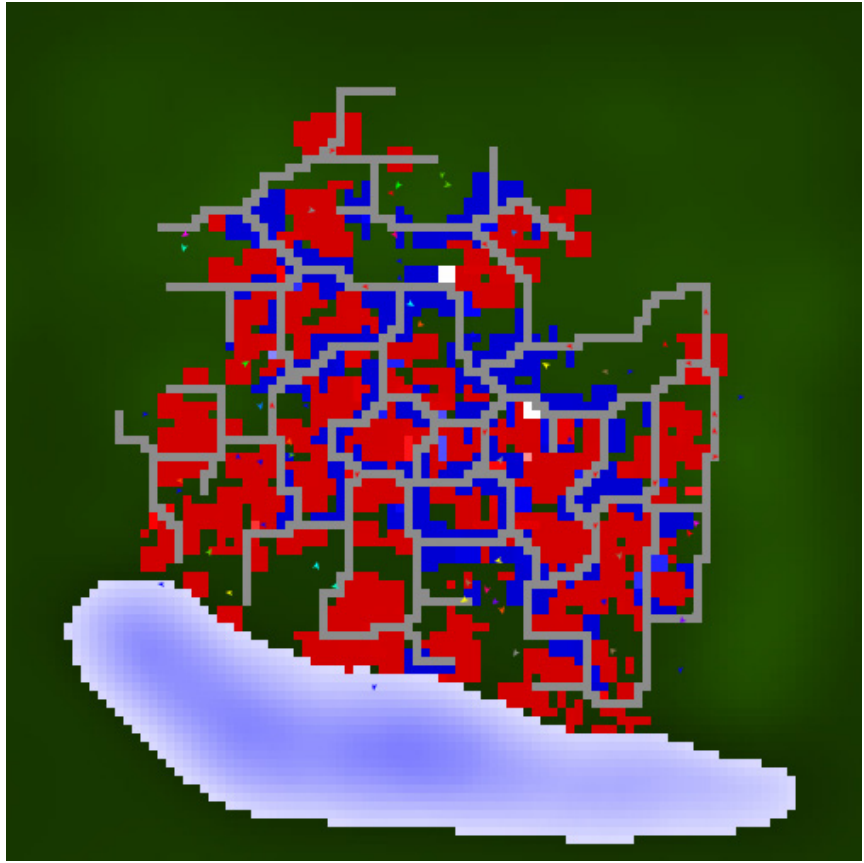

## Abstract

We propose a method to procedurally generate a familiar yet complex human artifact: the city. We are not trying to reproduce existing cities, but to generate artificial cities that are convincing and plausible by capturing developmental behavior. In addition, our results are meant to build upon themselves, such that they ought to look compelling at any point along the transition from village to metropolis. Our approach largely focuses upon land usage and building distribution for creating realistic city environments, whereas previous attempts at city modeling have mainly focused on populating road networks. Finally, we want our model to be self automated to the point that the only necessary input is a terrain description, but other high-level and low-level parameters can be specified to support artistic contributions.

With the aid of agent based simulation we are generating a system of agents and behaviors that interact with one another through their effects upon a simulated environment. Our philosophy is that as each agent follows a simple behavioral rule set, a more complex behavior will tend to emerge out of the interactions between the agents and their differing rule sets. By confining our model to a set of simple rules for each class of agents, we hope to make our model extendible not only in regard to the types of structures that are produced, but also in describing the social and cultural influences prevalent in all cities.

# Background

It has long been one of the aims of computer graphics to proceduralize models. As applications tend to grow in complexity and content, it becomes more and more arduous for artists and programmers to supply the necessary detail. In the past, computer graphics researchers have focused on automating models of natural phenomena, such as fire, clouds, water, and plants. Only recently has the computer graphics community focused its attention upon manmade artifacts.

The procedural generation of manmade artifacts such as cities is of particular relevance to the entertainment industry. With the latest trends in gaming hardware, gaming companies are placed under growing pressure to provide more content in their products, while still maintaining the same development cycle. This has already forced companies like Electronic Arts to hire an equal staff of artists and programmers. If circumstances were to continue as they are, such gaming companies would not be able to hire the necessary number of people to provide enough content by the shipping date. A tool that assisted the artist by proceduralizing the basic form of complex models while still allowing artistic contributions would be an immense asset.

Film companies have also shown great interest in procedural city modeling. Many times it is expensive for a company to shoot the camera in metropolises or to travel to foreign sites, and with the new ventures in special effects, more and more of the set has moved its way into the computer. Procedurally generated cities would supply directors with a low budget alternative with a wide spectrum of possible environments.

In addition to the entertainment industry, architects and urban planners also hold a great interest in this area of research. From an academic and industrial point of view, architects are constantly looking for innovative ways to display and present their ideas, not only to communicate to their students and clients, but also to assist in the creative process. Urban planners are also dependant on visualization, as they are constantly concerned with the implications that may come with their decisions. In fact, visualization of city data is an integral part of urban planning research.

Previous attempts of the graphics community to model cities have primarily focused upon road networks. The input to these methods tends to be a terrain specification and a population distribution from existing city data. Then an L-system grows the road network to service to the population density. Often the road networks have a structural design option, such as a gridded verses radial metric (Sun et al [2]). In the case of Parish et al [1], a height map was added as an extra constraint, and simple building geometry was generated as a function of the value of the population over a given plot. The shortcomings of these methods were their dependence on previously existing data, and the lack of blend between different road parameters. In the case of Parish et al, the automated architecture did not distinguish between building usage, resulting in whole cities that looked like downtown Manhattan.

Other research has been focused on the architecture of the city rather than its urban layout. Wonka et al [3] have come up with a method to generate detailed facades using a high level L-system. These buildings can be sensitive to cultural parameters as well as issues of population and construction materials. Yet even this method has its drawbacks due to its complexity. Since the model is based on an L-system, as different types of buildings are added, valid representations in the L-system encoding do not make sense when converted into 3D architecture. As a result, this method requires another set of constraints to monitor all the possible contradictions and prevent them from occurring. Such a solution rapidly grows in size and complexity with every building type that is added to the system. Finally, since their system does not account for an urban layout, they depend on detailed data from already existing cities, including building footprints as well as population densities.

# Our Model

## 1. Agent Based Simulation

Our model differs from the previous work mentioned in its usage of agents rather than an L-system. We realize that our model will need to have a large set of generic developers to generate the common components of the city (such as the residential, commercial, and industrial constituents), but it will need to support specialized agents to add the more refined and characterizing portions of the city (government buildings, squares, institutions). An L-system seemed inapt because of the parameter bloat that would result from all the specific exceptions and particularities that may come with a given culture. Agents and rule sets on the other hand seemed more appropriate, because it would be easier to describe how the agents ought to build their portion of the city rather than worry about what everything else is doing. Furthermore, the simpler the rules, the more robust the algorithm, and the less chance there will be for conflict.

Our agents interact directly with their environment, and only indirectly with each other, i.e. by observing the effects of other agents on the environment. This assumption in our model makes the construction of rules easier, because each agent only needs to concern itself with the local area around it, rather than consulting the global set of agents active in the world. To facilitate handling of the environment, the environment is represented as a rectangular set of patches. Like agents, each patch can also follow behaviors, but they are confined to always remain at the same point in space.

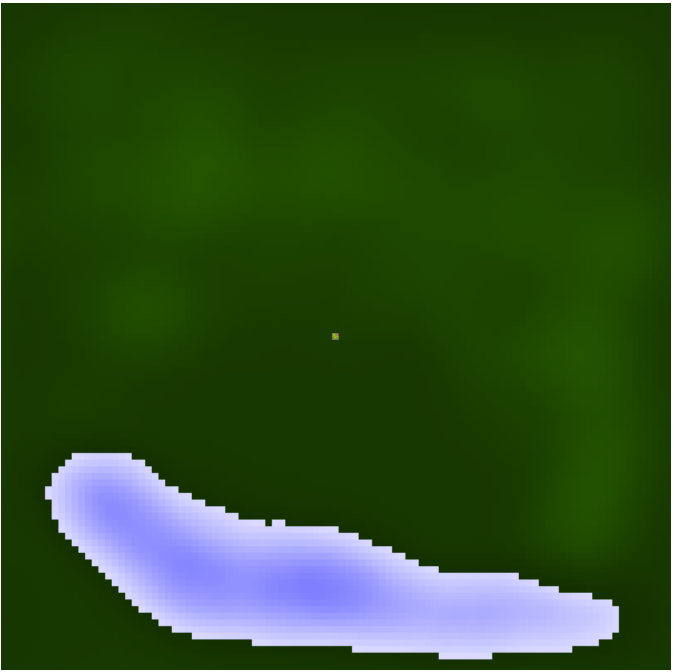

*Figure 1. A sample input file. Green depicts land; blue signifies water. Higher elevations are denoted by lighter shades, and lower elevations by darker. The spot in the middle is the seed that is used to start growing the city; it can be arbitrarily placed.*

## 2. Input

The core to the objective of our research is that we ought to be able to construct a unique, yet realistic city, within a given time and culture. With that having been said, the only low-level input necessary for our system is a terrain description. Terrain descriptions can be collected from existing sources, randomly generated, or artistically designed, but the time spent in constructing them is minimal, and the process itself is not complicated. Certain high-level parameters will also need to be set by the user. At the moment, we are only modeling Western culture within the last century. Once we have a working model with a set of parameters for this culture and era, we will construct packages of other rule sets and parameters.

Yet not all artists or architects may desire an entirely autonomous system. Many would rather see it as a tool instead of a possible replacement. Additionally, the interaction of human design with an intelligent model may be of more benefit than a simple file-in file-out operative. For this reason, we are extending our model to be able to specify attributes within a local context in the environment. An artist can choose which regions have what parameters to suit his or her own vision of how the city ought to develop. In short, the user ought to be able to specify as much or as little as he or she wants (within reason) in order to obtain a realistic looking city.

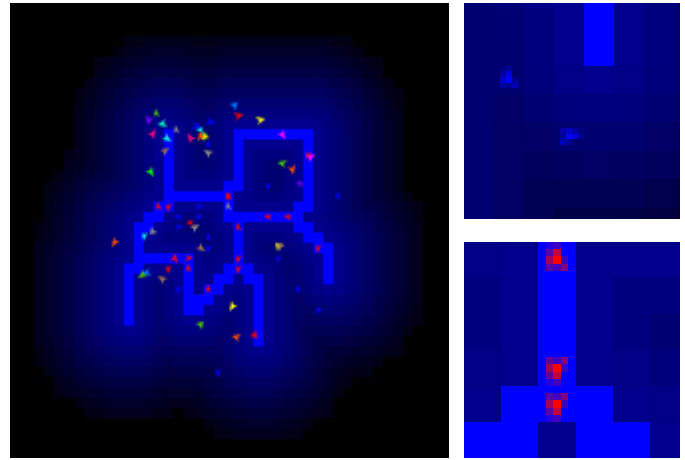

*Figure 2. Road generation. The light blue aura signifies the amount of influence the road has on the surrounding area. Top right: road extenders. Bottom right: road connectors.*

## 3. Road Generation

Our current version of our model only generates tertiary road networks. A tertiary road is a term used in urban design to specify the lowest level of road in a road hierarchy. Tertiary roads are often found within places such as residential neighborhoods and industrial parks. They are designed to service land within a certain distance, and they generally tend to be private, resulting in smaller road diameter and a variable level of interconnectivity.

The tertiary roads are generated through the efforts of two types of agents – extenders and connectors. When a road segment is laid down, it also stamps a distance value to each of the patches within a given radius. An extender agent roams the terrain until it finds a piece of land that is not serviced by the existing road network. Once that area of land has been discovered, it follows the distance values back to the road network, and keeps a record of all the patches it passes through during its journey. The terrain has some influence on the agent's decision, such that the agent tries to avoid taking a route with a large change in elevation. Once the agent reaches the network, it computes some constraint tests upon the connection. These tests involve issues like road density, proximity to existing intersections, and deviation from the starting position. If these tests pass, the road segment offered by the agent is confirmed and added to the existing road network and the extender continues to roam in search of more land to service. To prevent the road network from developing faster than the building construction, extenders are forbidden to roam too far from developed land.

A connector agent is constrained to wander over the existing road network. As it meanders, it samples a random patch in the road within a given radius. It then proceeds to try to reach the selected patch via the road network with a breadth first search. If it cannot reach its target within the original sampling radius, or if it must go too far out of its way, it will attempt to build a road segment between its current patch and the one it cannot reach. The suggested road segment is subjected to the same constraint tests as those supplied by the extender agents before it is added to the network.

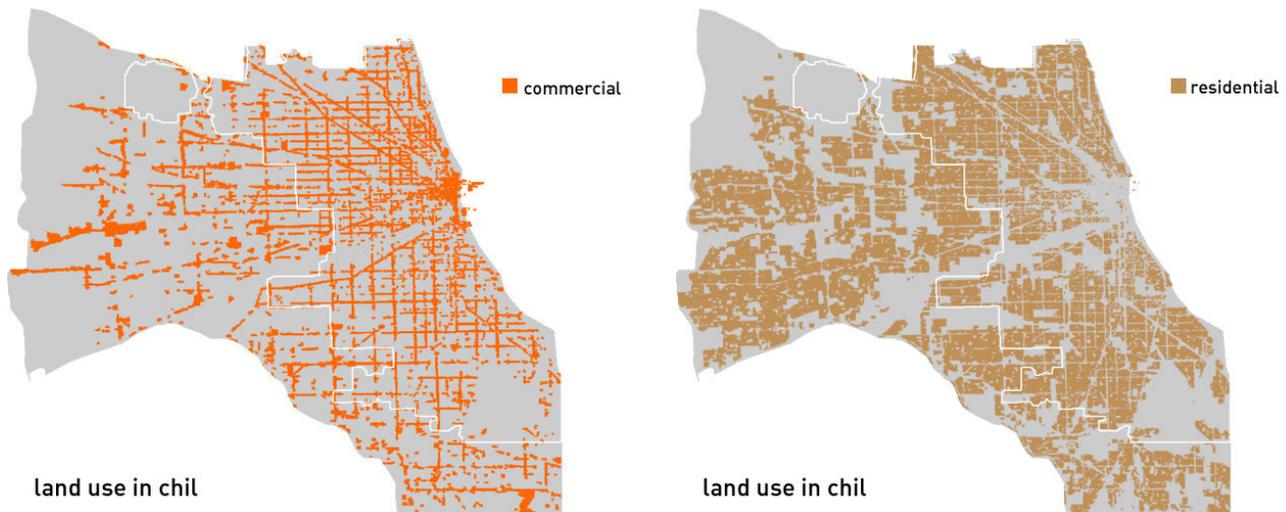

*Figure 3. Commercial and residential land usage distributions in modern Chicago, IL. Note how commercial zones tend to cluster around the road network, while residents prefer to spread out in regions of lesser road density. Courtesy of UrbanLab.*

The constraints imposed on our road-building agents can be used to simulate a range of differing behaviors. The user can specify that the road segments must lie directly on a pre-established gridded pattern, or give the road segments permission to be located within a certain deviation from the grid. When the deviation is set very large in proportion to the length of the submitted road segments, it has the effect of looking totally organic. As this deviation is set towards zero, the roads become more rectilinear.

Another set of constraints limits the interconnectivity of the road network. Road connectors will only add a road segment if it takes longer than some constant *k* times to reach the target than the Manhattan-distance from the starting position to the target. If this constant is set low, the road network becomes tighter with more loops. If it is set higher, the road network has more dead ends, with a more private atmosphere.

## 4. Building Developers

Ultimately, our model will include at least one developer for every major land use in the city. Specific subclasses of these main developer types may be introduced later in order to model particular cultural nuances. At this moment, only two building developers have been implemented: commercial and residential developers. So far, their behavior is identical, except that commercial developers prefer to build on land with high road density within close distance to the roads themselves, while residential developers tend to shy away from these areas (see Figure 3). Both types of agents like to cluster with each other and try to avoid disparities in population density. Land that is near water and land that has an elevation slightly above the average city elevation is considered to be more valuable.

Each developer roams the world within a certain distance of the road network. When a developer lands on an empty patch, it evaluates the value of that patch. Then it suggests a building plot for that site, which can consist of multiple patches, and evaluates whether the new building raises the value of the suggested plot. If the value is increased, the building is committed to the site, otherwise the developer moves on to consider another site. Should the developer happen to land on a site of the same building type it can produce, it will raises the population density of the building if it results in an increase in value.

## 5. Near Future Goals

We have a number of improvements and components to add to our model that we plan to include in the next couple of weeks.

First, we plan on integrating another level of road networks into the road hierarchy. This level would be used to model primary roads. A primary road holds higher importance in the road network. They tend to follow territorial boundaries and provide flow to separate areas of the city, rather than directly servicing the immediate land around them.

Second, we will introduce "speculators" which will behave like "pseudo-developers." Instead of developing the land directly, they will merely go about reserving large tracts of land to be sold off to developers at a later time. In this way, we hope to simulate more high-level urban planning, such as residential districts, universities, and parks.

Third, our plot formation for individual buildings is still in its primitive form. In our current models, plots still tend to be single patches, and only occasionally span a few more than that. We believe that the addition of multiple-patched plots will make our results look more convincing.

Fourth, we have established collaboration with EA, who will be allowing us to utilize their SimCity software to visualize our results. We plan to map our results to their models in a post process.

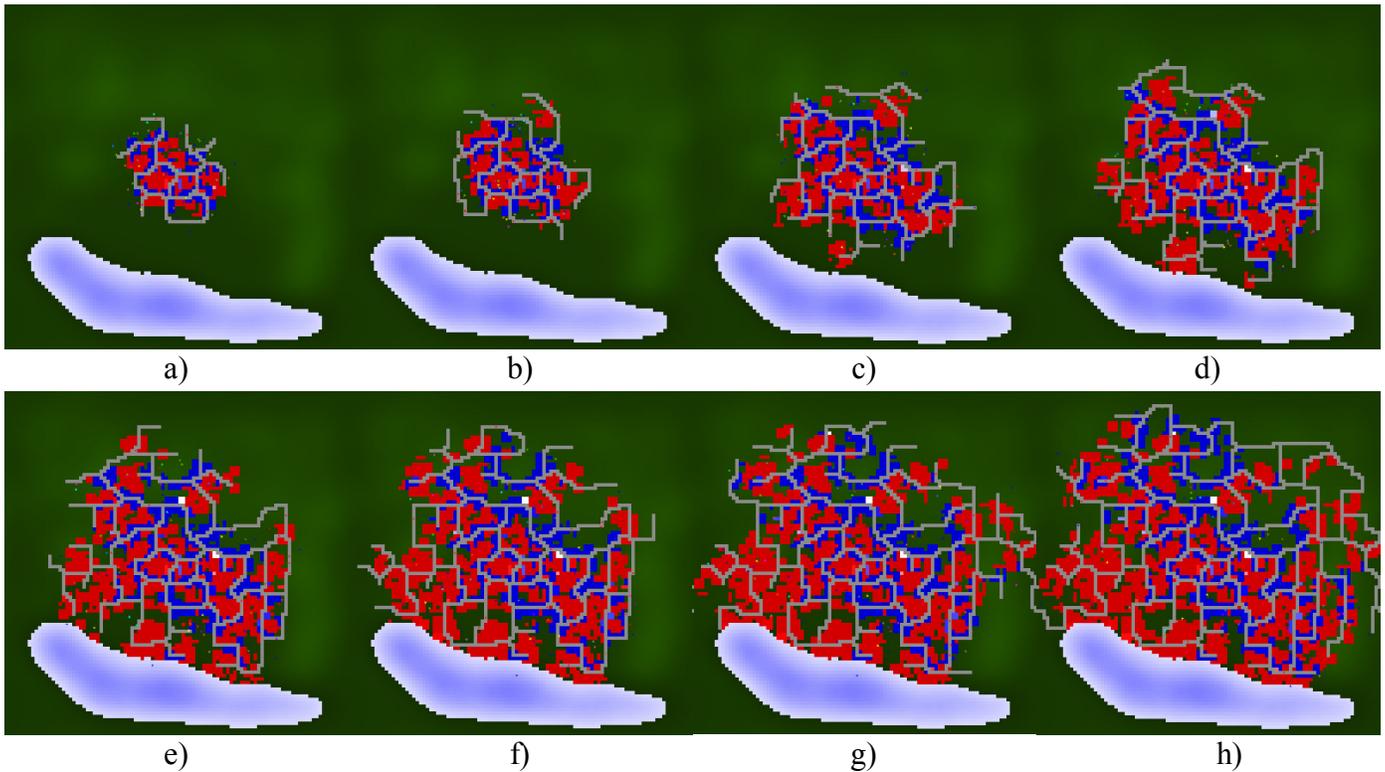

Figure 4. A progression of results taken at subsequent intervals during a simulation of an organic city.

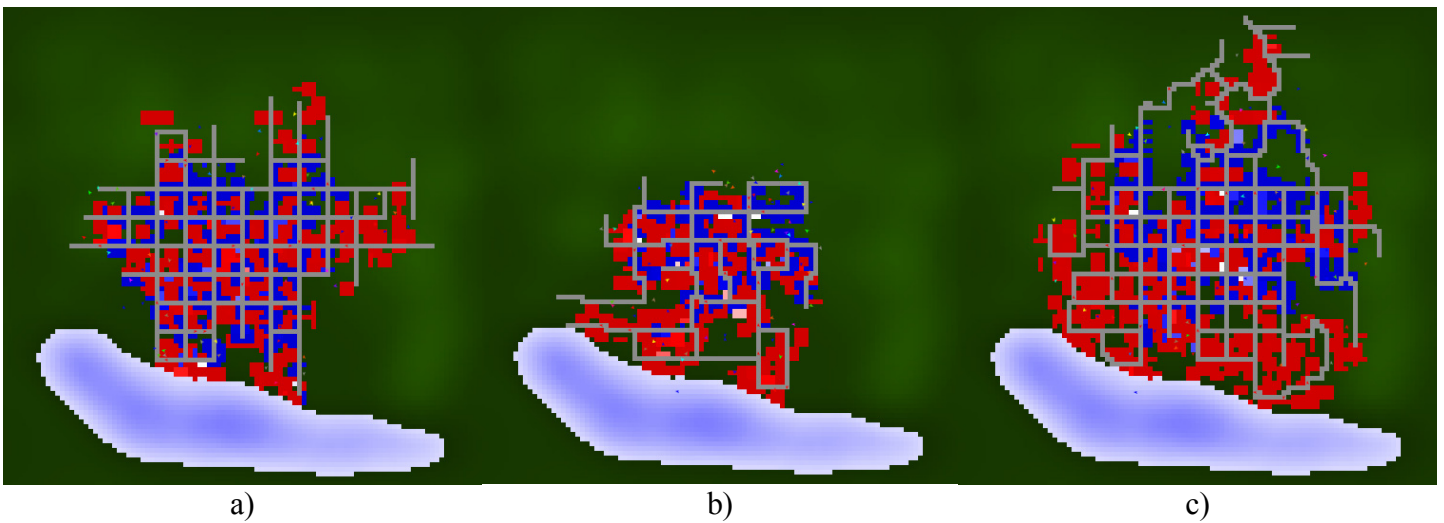

Figure 5: Examples of differing city structures based on road network strategies: a) tightly gridded, b) lax grid with loose connectivity, c) mixture of gridded and organic behaviors.

## Results

Our model has been implemented in NetLogo, an agent based programming environment written in Java under the leadership of Uri Wilensky. In general, our simulations run in real time, and users can tweak parameters and watch the city build before their eyes. The results in Figure 4 were taken from the same simulation, which ran for about half an hour. The results in Figure 5 were each separate simulations with different city attributes. Each ran for about fifteen minutes.

The blue regions represent commercial development and the red represent residential. Lighter shades of color signify higher density - whitish areas are those of the highest density. The road networks are represented in gray. It is interesting to see how the simple variation in road structure can dramatically alter the city representation. Note that even with this early approach at developing commercial and residential behavior we can observe similar results to what has been obtained in the Chil data (Figure 3). The commercial areas are concentrated, stay closer to the road network, and remain within the more developed parts of the city. In contrast, the residential developers are more pioneering and tend to spread out in areas of lesser road density.

# Future Work

We believe that our model is showing much progress towards generating a realistic city environment, but we realize that there is still much to be done before its results look truly accurate. In our future research we plan to add agents to build higher networks of transportation, such as highways and public transportation systems. We also will add agents to satisfy the other major land usage categories, including government buildings, cemeteries, and industry. As our model for Western cultures becomes more complete, we will expand our research to support other urban layouts from different cultures and historical periods. Finally, we plan to incorporate our output into our own 3D architectural generator to populate our virtual cities instead of relying on commercial software.

# Acknowledgements

We would like to acknowledge the following for their indispensable contribution and support of this project: the National Science Foundation, for their generous funding support; Electronic Arts, for their valuable suggestions about user interactivity, their generous software contribution, SimCity 3000, and their gracious hospitality; SimCenter, for their mentoring and advice concerning urban planning; and Peter Wonka and Bill Ribarsky (authors of Instant Architecture, SIGGRAPH '03), for their insights and collaboration.